\documentclass[12pt]{article}
\usepackage[english,german,french,polish]{babel}
\usepackage[T1]{fontenc}
\usepackage{amsfonts}

\begin{document}

\selectlanguage{english}

\baselineskip 0.76cm
\topmargin -0.6in
\oddsidemargin -0.1in

\let\ni=\noindent

\renewcommand{\thefootnote}{\fnsymbol{footnote}}

\pagestyle {plain}

\setcounter{page}{1}

\pagestyle{empty}

~~~

\begin{flushright}
IFT--04/27
\end{flushright}

{\large\centerline{\bf The four-group $Z_2\times Z_2$ as a discrete invariance group}}

{\large\centerline{\bf of effective neutrino mass matrix{\footnote {Work supported in part by the Polish State Committee for Scientific Research (KBN), grant 2 P03B 129 24 (2003--2004).}}}}

\vspace{0.4cm}

{\centerline {\sc Wojciech Kr\'{o}likowski}}

\vspace{0.3cm}

{\centerline {\it Institute of Theoretical Physics, Warsaw University}}

{\centerline {\it Ho\.{z}a 69,~~PL--00--681 Warszawa, ~Poland}}

\vspace{0.6cm}

{\centerline{\bf Abstract}}

\vspace{0.2cm}

Two sets of four $3\times 3$ matrices $ {\bf 1}^{(3)},\, \varphi_1, \,\varphi_2, \,\varphi_3$ and  ${\bf 1}^{(3)}, \,\mu_1, \,\mu_2, \,\mu_3$ are constructed, forming two unitarily isomorphic reducible representations $\underline{3}$ of the group $Z_2\times Z_2$ called often the four-group. They are related to each other through the effective neutrino mixing matrix $U$ with $s_{13} = 0$, and generate four discrete transformations of flavor and mass active neutrinos, respectively. If and only if $s_{13} = 0$, the generic form of effective neutrino mass matrix $M$ becomes invariant under the subgroup $Z_2$ of $Z_2\times Z_2$ represented by the matrices $ {\bf 1}^{(3)}$ and $\varphi_3$. In the approximation of $m_1 = m_2$, the matrix $M$ becomes invariant under the whole $Z_2\times Z_2$ represented by the matrices  $ {\bf 1}^{(3)}, \,\varphi_1, \,\varphi_2, \,\varphi_3$. The effective neutrino mixing matrix $U$ with $s_{13} = 0$ is always invariant under the whole $Z_2\times Z_2$ represented in two ways, by the matrices $ {\bf 1}^{(3)}, \,\varphi_1, \,\varphi_2, \,\varphi_3$ and $ {\bf 1}^{(3)}, \,\mu_1, \,\mu_2, \,\mu_3$. 

\vspace{0.2cm}

\ni PACS numbers: 12.15.Ff , 14.60.Pq , 12.15.Hh .

\vspace{0.6cm}

\ni October 2004 

\vfill\eject

~~~
\pagestyle {plain}

\setcounter{page}{1}

\vspace{0.3cm}

\ni {\bf 1. Introduction}

\vspace{0.2cm}

As is well known, the neutrino experiments with solar  $\nu_e$'s [1], atmospheric $\nu_\mu$'s [2], long-baseline accelerator $\nu_\mu$'s [3] and long-baseline reactor $\bar{\nu}_e$'s [4] are very well described by oscillations of three active neutrinos $\nu_e \,,\, \nu_\mu \,,\, \nu_\tau $, where the mass-squared splittings of the related neutrino mass states $\nu_1\,,\, \nu_2 \,,\, \nu_3 $ are estimated to be $\Delta m^2_{\rm sol} \equiv \Delta m^2_{21} \sim 8\times 10^{-5}\; {\rm eV}^2$ and $\Delta m^2_{\rm atm} \equiv \Delta m^2_{32} \sim 2.5\times 10^{-3}\; {\rm eV}^2$ [5]. The effective neutrino mixing matrix $U = \left(U_{\alpha i} \right) \;(\alpha = e, \mu, \tau\;{\rm and}\; i=1, 2, 3)$, responsible for the unitary transformation

\begin{equation}
\nu_\alpha  = \sum_i U_{\alpha i}\, \nu_i \;,
\end{equation}

\ni is experimentally consistent with the global bilarge form

\begin{equation}
U = \left( \begin{array}{ccc} c_{12} & s_{12} & 0 \\ - c_{23} s_{12} & c_{23} c_{12} & s_{23} \\ s_{23} s_{12} & -s_{23} c_{12} & c_{23}  \end{array} \right) \;,
\end{equation}

\vspace{0.2cm}

\ni where $c_{ij} = \cos \theta_{ij}$ and $s_{ij} = \sin \theta_{ij}$ with the estimations $\theta_{23} \sim 45^\circ $ and $\theta_{12} \sim 32^\circ $ ({\it i.e.}, $c_{23} \sim 1/\sqrt2 \sim s_{23}$),  while the matrix element $U_{e3} = s_{13} \exp(-i\delta)$ is neglected due to the nonobservation of neutrino oscillations for short-baseline reactor $\bar{\nu}_e$'s, especially in the Chooz experiment [6] giving for  $s^2_{13}$ the upper limit $s^2_{13} < 0.04$. We assume here that $0 \leq \theta_{13} \leq \pi/2$, thus $s_{13} = 0$ implies $c_{13} = 1$.

However, the mixing matrix (1) (involving two experimentally fitted mass-squared scales $\Delta m^2_{21}$ and $\Delta m^2_{32}$) cannot explain the possible LSND effect for short-baseline accelerator $\bar{\nu}_\mu$'s [7] that should require the existence of a third independent neutrino mass-squared scale, say, $\Delta m^2_{\rm LSND} \sim 1\; {\rm eV}^2$. Unless the CPT invariance is seriously violated in neutrino oscillations [8] (leading to considerable mass splittings between neutrinos and antineutrinos), such a third scale cannot appear in the oscillations of three neutrinos.
So, if the ongoing MiniBooNE experiment [9] confirmed the LSND result, we should need one, at least, light sterile neutrino in addition to three active neutrinos in order to introduce the third scale (in the case, when the serious CPT violatrion was excluded).

The effective neutrino mass matrix $ M = \left( M_{\alpha \beta}\right)\;(\alpha, \beta = e , \mu , \tau )$ is connected with the neutrino mixing matrix $U$ through the formula:

\begin{equation}
M_{\alpha \beta}  = \sum_i U_{\alpha i}\, m_i\, U^*_{\beta i}\,,
\end{equation}

\ni if the flavor representation is used, where the mass matrix of charged leptons $e^- , \mu^- , \tau^- $  is diagonal and so, the mixing matrix $U$ is at the same time the diagonalizing matrix for $M$, 

\begin{equation}
 \sum_{\alpha, \beta} U^*_{\alpha i} M_{\alpha\,\beta}  U_{\beta j} = m_i \delta_{ij}
\end{equation}

\ni (we assume that $ M^\dagger = M^* = M$ for simplicity). Applying the generic form of the effective neutrino mixing matrix 

\begin{equation}
U = \left(\begin{array}{ccc} 1 & 0 & 0 \\ 0 & c_{23} & s_{23} \\ 0 & -s_{23} & c_{23} \end{array}\right)
\left(\begin{array}{ccc} c_{13} & 0 & s_{13} \\ 0 & 1 & 0 \\ -s_{13} & 0 & c_{13} \end{array}\right) \left(\begin{array}{ccc} c_{12} & s_{12} & 0 \\ -s_{12} & c_{12} & 0 \\ 0 & 0 & 1 \end{array}\right) 
\end{equation}

\vspace{0.2cm}

\ni (without one Dirac and two Majorana CP-violating phases for simplicity) that is reduced to the form (2) if $s_{13} = 0$, we obtain from Eq. (3)

\begin{eqnarray}
M_{e e} & = & c^2_{13}\left(c^2_{12} m_1 + s^2_{12} m_2\right) + s^2_{13} m_3 \,, \nonumber \\ 
M_{\mu\,\mu} & = & c^2_{23}\left(s^2_{12} m_1 + c^2_{12} m_2\right) + s^2_{23} \left[ s^2_{13}\left(c^2_{12} m_1 + s^2_{12} m_2\right) + c^2_{13} m_3\right]  \nonumber \\ 
& & \!\!\!+\, 2c_{23}s_{23} s_{13} c_{12} s_{12} (m_1 - m_2) \,, \nonumber \\
M_{\tau\,\tau} & = & s^2_{23}\left(s^2_{12} m_1 + c^2_{12} m_2\right) + c^2_{23}\left[ s^2_{13}\left(c^2_{12} m_1+ s^2_{12} m_2\right) + c^2_{13} m_3\right] \nonumber \\
& &  \!\!\! - 2c_{23}s_{23} s_{13} c_{12} s_{12} (m_1 - m_2)  \,, \nonumber \\ 
M_{e\,\mu } & = & \!\!\!- c_{23} c_{13} c_{12} s_{12}(m_1 - m_2) - s_{23} c_{13} s_{13} \left(c^2_{12}m_1 + s^2_{12} m_2 - m_3\right)  \,, \nonumber \\
M_{e\,\tau } & = &  s_{23} c_{13} c_{12} s_{12}(m_1 - m_2) - c_{23} c_{13} s_{13} \left(c^2_{12}m_1 + s^2_{12} m_2 - m_3\right)  \,, \nonumber \\
M_{\mu\,\tau } & = & \!\!\!- c_{23} s_{23}\left[s^2_{12} m_1 + c^2_{12} m_2 - s^2_{13}\left(c^2_{12} m_1 + s^2_{12} m_2\right) - c^2_{13} m_3 \right] \nonumber \\
& & \!\!\!+\, c_{12}s_{12} s_{13}\left(c^2_{23} - s^2_{23} \right) (m_1 - m_2) \,.
\end{eqnarray}

\ni If $s_{13} = 0$, Eqs. (6) can be rewritten in the matrix form as follows:

\begin{eqnarray}
M & = & \;\,\frac{m_1+m_2}{2} \left( \begin{array}{rrr} 1 & 0 & 0 \\ 0 & c^2_{23} & -c_{23} s_{23} \\ 0 & -c_{23} s_{23} & s^2_{23} \end{array} \right) + m_3 \left( \begin{array}{rrr} 0 & 0 & 0 \\ 0 & s^2_{23} & c_{23} s_{23}  \\ 0 & c_{23} s_{23}  & c^2_{23} \end{array}\right) \nonumber \\ & & \!\!\! + \,\frac{m_1-m_2}{2} \left( \begin{array}{rrr} c_{\rm sol} & -c_{23} s_{\rm sol}  & s_{23}s_{\rm sol} \\ -c_{23} s_{\rm sol} & -c^2_{23} c_{\rm sol} & c_{23} s_{23} c_{\rm sol} \\ s_{23} s_{\rm sol} & c_{23} s_{23} c_{\rm sol} & -s^2_{23}c_{\rm sol} \end{array}\right) \,,
\end{eqnarray}

\ni where $c_{\rm sol} = c^2_{12} - s^2_{12} = \cos 2 \theta_{12}$ and $s_{\rm sol} = 2c_{12} s_{12} = \sin 2 \theta_{12}$. In Eq. (7), all three $3\times 3$ matrices on its rhs commute, while its third term is of the form

\begin{equation}
 \frac{m_1-m_2}{2} \left[c_{\rm sol} \left( \begin{array}{rrr} 1 & 0 & 0 \\ 0 & -c^2_{23} & c_{23} s_{23} \\ 0 & c_{23} s_{23} & -s^2_{23} \end{array} \right) + s_{\rm sol} \left( \begin{array}{rrr} 0 & -c_{23} & s_{23} \\ -c_{23} & 0 & 0  \\ s_{23} & 0 & 0 \end{array}\right) \right]
\end{equation}

\vspace{0.2cm}

\ni involving two anticommuting $3\times 3$ matrices. Diagonalizing both sides of Eq. (7), one gets consistently

\begin{eqnarray}
\left( \begin{array}{rrr} m_1 & 0 & 0 \\ 0 & m_2 & 0 \\ 0 & 0 & m_3 \end{array} \right) & = & \frac{m_1 + m_2}{2} \left( \begin{array}{rrr} 1 & 0 & 0 \\ 0 & 1 & 0 \\ 0 & 0 & 0 \end{array} \right) + \,m_3 \left( \begin{array}{rrr} 0 & 0 & 0 \\ 0 & 0 & 0 \\ 0 & 0 & 1 \end{array} \right) \nonumber \\ &+ & \frac{m_1 - m_2}{2} \left( \begin{array}{rrr} \!\!1 & 0 & 0 \\ \!\!0 & -1 & 0 \\ \!\!0 & 0 & 0 \end{array} \right) \,. 
\end{eqnarray}

\vspace{0.3cm}

\ni {\bf 2. Invariance of effective mass matrix $M$ in the case of $s_{13} = 0$}

\vspace{0.2cm}

Introduce three discrete transformations of active neutrinos $\nu_e ,\, \nu_\mu ,\, \nu_\tau, $

\begin{equation}
\left( \begin{array}{r} \nu'_e \\ \nu'_\mu \\ \nu'_\tau \end{array} \right)_{\!\!a} = \varphi_a \left( \begin{array}{r} \nu_e \\ \nu_\mu \\ \nu_\tau \end{array} \right) \;\;(a = 1,2,3) ,
\end{equation}

\ni where

\begin{eqnarray} 
\varphi_1 & \equiv & \left( \begin{array}{rrr} -1 & 0 & 0 \\ 0 & c_{\rm atm} & -s_{\rm atm} \\ 0 & -s_{\rm atm} & -c_{\rm atm} \end{array} \right) = {\rm diag}\left(-1,c_{\rm atm}\sigma_3 - s_{\rm atm} \sigma_1\right)  \!\stackrel{s_{\rm atm} \!\rightarrow \pm 1}{\longrightarrow}\!\!  \left( \begin{array}{rrr} -1 & 0 & 0 \\ 0 & 0 & \mp 1 \\ 0 & \mp1 & 0 \end{array} \right) \,, \nonumber \\
\varphi_2 & \equiv & \left( \begin{array}{rrr} 1 & 0 & 0 \\ 0 & -1 & 0 \\ 0 & 0 & -1 \end{array} \right) = {\rm diag}\left(1\,,\, -{\bf 1}^{(2)}\right) \,, \nonumber  \\
\varphi_3 & \equiv & \left( \begin{array}{rrr} -1 & 0 & 0 \\ 0 & -c_{\rm atm} & s_{\rm atm} \\ 0 & s_{\rm atm} & c_{\rm atm} \end{array} \right)= {\rm diag}\left(-1,-c_{\rm atm}\sigma_3 \!+\! s_{\rm atm} \sigma_1\right) \! \stackrel{s_{\rm atm} \rightarrow \pm 1}{\longrightarrow}\! \left( \begin{array}{rrr} -1 & 0 & 0 \\ 0 & 0 & \pm 1 \\ 0 & \pm 1 & 0 \end{array} \right) \nonumber \\ & &
\end{eqnarray}

\vspace{0.2cm}

\ni are $3\times 3$ Hermitian matrices with $c_{\rm atm} = c^2_{23} - s^2_{23} = \cos 2\theta_{23} $ and $s_{\rm atm} = 2c_{23} s_{23} = \sin 2\theta_{23} $ (here, the approximation of $s_{\rm atm} = \pm 1$ {\it i.e.}, $c_{23} = 1/\sqrt2 = \pm s_{23}$ is experimentally satisfactory). Similarly, the $3\times 3$ unit matrix  

\begin{equation}
{\bf 1}^{(3)} \equiv \left( \begin{array}{rrr} 1 & 0 & 0 \\ 0 & 1 & 0 \\ 0 & 0 & 1 \end{array} \right) = {\rm diag}(1, {\bf 1}^{(2)})\,,
\end{equation}

\vspace{0.2cm}

\ni where ${\bf 1}^{(2)} = {\rm diag}(1, 1)$, describes the active-neutrino identity transformation. The matrices (11) were already used in Ref. [10], but in the limit of $s_{\rm atm} \rightarrow 1$.

It is easy to see that the four matrices ${\bf 1}^{(3)},\, \varphi_1 \,,\,\varphi_2 \,,\, \varphi_3$ satisfy for any $c_{\rm atm}$ and $s_{\rm atm}$ the following algebraic relations

\begin{equation}
\varphi_1 \varphi_2 = \varphi_3\;\,{\rm (cyclic)} \;\,,\;\, \varphi^2_a= {\bf 1}^{(3)}  \;\,,\;\, \varphi_a \varphi_b = \varphi_b \varphi_a 
\end{equation}

\ni and also the constraint


\begin{equation}
{\bf 1}^{(3)} + \varphi_1+ \varphi_2 + \varphi_3 = 0 \,.
\end{equation}

\ni The Cayley table equivalent to the relations (13) gets the form 

\begin{center}
\begin{tabular}{l|llll}
 & ${\bf 1}^{(3)}$ & $\varphi_1$ & $\varphi_2$ & $\varphi_3$ \\ \hline
${\bf 1}^{(3)}$ & ${\bf 1}^{(3)}$ & $\varphi_1$ & $\varphi_2$ & $\varphi_3$ \\
$\varphi_1$ & $\varphi_1$ & ${\bf 1}^{(3)}$ & $\varphi_3$ & $\varphi_2$ \\ 
$\varphi_2$ & $\varphi_2$ & $\varphi_3$ & ${\bf 1}^{(3)}$ & $\varphi_1$ \\ 
$\varphi_3$ & $\varphi_3$ & $\varphi_2$ & $\varphi_1$ & ${\bf 1}^{(3)}$ \\
\end{tabular}
\end{center}

The algebraic relations (13) with ${\bf 1}^{(3)}$ replaced by the generic unit element {\bf 1} (and $\varphi_1$, $\varphi_2$, $\varphi_3$ --- by three other generic group elements) characterize a finite group $Z_2\times Z_2$ of the order four often called the four-group [11] ($Z_2$ is the cyclic group of the order two). It is isomorphic to the dihedral group [11] of the order four and also to the group of four special permutations

\begin{equation}
\left(\begin{array}{rrrr} 1 & 2 & 3 & 4 \\ 1 & 2 & 3 & 4 \end{array}\right) \;,\; \left(\begin{array}{rrrr} 1 & 2 & 3 & 4 \\ 2 & 1 & 4 & 3 \end{array}\right) \;,\;\left(\begin{array}{rrrr} 1 & 2 & 3 & 4 \\ 3 & 4 & 1 & 2 \end{array}\right) \;,\; \left(\begin{array}{rrrr} 1 & 2 & 3 & 4 \\ 4 & 3 & 2 & 1 \end{array}\right) 
\end{equation}

\vspace{0.2cm}

\ni of four objects. As is known, all final groups of the order four are isomorphic either to the four-group $Z_2\times Z_2$ or to the cyclic group $Z_4$ of the order four, these two being not isomorphic to each other. Both are Abelian. Note that the dihedral group of the order six is isomorphic to the permutation group $S_3$ of three objects, and the dihedral group of the order eight is the group $D_4$ considered in Refs. [12] and [13]. They are non-Abelian.

Four $3\times 3$ matrices (12) and (11) constitute a reducible representation $\underline{3} = \underline{1} + \underline{2}$ of the four-group, where its  representations $\underline{1}$ and $\underline{2}$ consist of four numbers

\begin{equation}
1\;,\;-1\;,\;1\;,\;-1
\end{equation}

\ni and of four $2\times 2$ matrices

\begin{equation}
{\bf 1}^{(2)}\;,\; c_{\rm atm} \sigma_3 - s_{\rm atm} \sigma_1\;,\;-{\bf 1}^{(2)}\;,\; -c_{\rm atm} \sigma_3 + s_{\rm atm} \sigma_1\,,
\end{equation}

\ni respectively. The representations $\underline{1}$ and $\underline{2}$ are, respectively, irreducible and reducible but the second is {\it not} reduced (to the sum diag$(\underline{1}, \underline{1})$ of two irreducible representations $\underline{1}$ consisting of four numbers $1,1,-1,-1$ and $1,-1,-1,1$). The constraint (14) with ${\bf 1}^{(3)}$ replaced by the generic unit element {\bf 1} (and $\varphi_1,\varphi_2,\varphi_3$ --- by three other generic group elements) is satisfied in both cases. But this constraint is not included in the definition of the four-group.

The four permutations (15) can be represented as a reducible representation  $\underline{4} = \underline{2} + \underline{2}$ of the four-group consisting of four $4\times 4$ matrices ${\bf 1}^{(D)} ,\, \sigma^{(D)}_1,\, \gamma_5\,,\, \gamma_5 \sigma^{(D)}_1 = \alpha_1$, where ${\bf 1}^{(D)} ,\, \sigma^{(D)}_1 ,\, \gamma_5$ are formal Dirac $4\times 4$ matrices in the Dirac representation: $\, \sigma^{(D)}_1= $ diag$(\sigma_1,\sigma_1)$,  $ \gamma_5 = $ antidiag(${\bf 1}^{(2)} ,\, {\bf 1}^{(2)}$) and, as always, ${\bf 1}^{(D)} = $ diag (${\bf 1}^{(2)}, {\bf 1}^{(2)}$). After a unitary transformation of Dirac matrices, one can write $ \gamma_5 = $ diag(${\bf 1}^{(2)} \,,\, -{\bf 1}^{(2)}$) and still $ \sigma^{(D)}_1= $ diag$ (\sigma_1, \sigma_1) $, as in the chiral representation. Then, $\underline{4} = $ diag $(\underline{2}, \underline{2})$, where the second of two not reduced representations $\underline{2} $ of the four-group  is identical with its representation (17), when $c_{\rm atm} = 0$ and $s_{\rm atm} = -1 $ (or $s_{\rm atm} = 1$, but then the elements $\,\sigma^{(D)}_1 $ and $\gamma_5 \sigma^{(D)}_1$ of $\underline{4}$ are interchanged). Putting formally $(1,2,3,4)^T = (-\nu_e/\sqrt2 \,,\,\nu_e/\sqrt2 \,,\,\nu_\mu\,,\,\nu_\tau)^T$,  one would obtain from the reducible representation $\underline{4}$ of the four-group its reducible representation $ \underline{1} + \underline{1} + \underline{2} = \underline{1} + \underline{3} $, where $ \underline{1} = (1,-1,1,-1)$ and $\underline{3} = ({\bf 1}^{(3)},\, \varphi_1\,,\, \varphi_2\,,\, \varphi_3)$ as given in Eqs. (16) and (12), (11) with $c_{\rm atm} = 0$ and $s_{\rm atm} = -1 $ (or $s_{\rm atm} = 1$, but then the elements $\varphi_1 $ and $\varphi_3 $ of $\underline{3}$ are interchanged). In general, the representation $\underline{4}$ of the four-group might suggest the existence of a fourth light neutrino, $\nu_s$, sterile in the gauge interactions of Standard Model. Then, the entries 1 and 2 in $(1,2,3,4)^T$ could be expressed through $\nu_s$ and $\nu_e$. Note, however, that even in this case the strile neutrino $\nu_s$ may be eliminated, if the constraint

$$
\left({\bf 1}^{(D)} + \sigma_1^{(D)} + \gamma_5 + \gamma_5 \sigma_1^{(D)} \right)(1,2,3,4)^T = 0
$$

\ni is imposed on $(1,2,3,4)^T$. In fact, with $\gamma_5 = $ diag$({\bf 1}^{(2)},-{\bf 1}^{(2)})$ this constraint is split into two conditions

$$
\left({\bf 1}^{(2)} + \sigma_1 + {\bf 1}^{(2)} + \sigma_1 \right)(1,2)^T = 0\;\,{\rm and}\;\,\left({\bf 1}^{(2)} + \sigma_1 - {\bf 1}^{(2)} - \sigma_1 \right)(3,4)^T = 0
$$

\ni of which the second is satisfied for 3 and 4 identically, while the first with $\sigma_1 = {\rm  antidiag}(1,1)$ gives for 1 and 2 one condition (1) + (2) = 0 implying that from two orthogonal superpositions ${\footnotesize \frac{1}{\sqrt2}}\left[(1)+(2) \right]$ and ${\footnotesize \frac{1}{\sqrt2}}\left[(2)-(1) \right]$ only the second survives. Then, identifying $\nu_s$ and $\nu_e$ with the first and the second superposition, respectively, one obtains $\nu_s = 0$ and $\nu_e = (2) \sqrt2 = -(1)\sqrt2$. In such a way $\nu_s$ may be eliminated indeed, giving $(1,2,3,4)^T = (-\nu_e/\sqrt2, \nu_e/\sqrt2, \nu_\mu, \nu_\tau)^T$, just as was formally considered above. After a unitary transformation, where $(1) \rightarrow {\footnotesize \frac{1}{\sqrt2}}\left[(1)+(2) \right] = 0$ and $(2) \rightarrow {\footnotesize \frac{1}{\sqrt2}}\left[(2)-(1) \right] = (2) \sqrt2$, one gets $(-\nu_e/\sqrt2, \nu_e/\sqrt2, \nu_\mu, \nu_\tau)^T \rightarrow (0, \nu_e, \nu_\mu, \nu_\tau)^T $. Notice that, after this transformation, the first of two representations  $ \underline{2}$ in $\underline{4} = {\rm diag}(\underline{2}, \underline{2})$ becomes reduced to the sum diag$(\underline{1}, \underline{1})$ of two irreducible representations $\underline{1}$ consisting of four numbers $1,1,1,1$ and $1,-1,1,-1$.

Making use of the matrices ${\bf 1}^{(3)},\,\varphi_1\,,\,\varphi_2\,,\, \varphi_3$ as given in Eqs. (12) and (11), we can rewrite the formula (7) for the effective neutrino mass matrix, valid in the case of $s_{13} = 0$, as follows:

\begin{eqnarray}
M & = & \;\,\frac{m_1+m_2}{2}\frac{1}{2}\left({\bf 1}^{(3)} - \varphi_3\right) + m_3 \frac{1}{2}\left({\bf 1}^{(3)} + \varphi_3\right)  
\nonumber \\ 
& & - \frac{m_1-m_2}{2} \left[ c_{\rm sol} \frac{1}{2}\left( \varphi_1 - \varphi_2 \right) + s_{\rm sol}\frac{1}{2}\left(c_{23}\lambda_1 - s_{23}\lambda_4 \right)\right]\,,
\end{eqnarray}

\ni where

\begin{equation}
\lambda_1 = \left( \begin{array}{rrr} 0 & 1 & 0 \\ 1 & 0 & 0 \\ 0 & 0 & 0 \end{array} \right) \;,\;
\lambda_4 = \left( \begin{array}{rrr}  0 & 0 & 1 \\ 0 & 0 & 0 \\ 1 & 0 & 0 \end{array} \right) 
\end{equation}

\vspace{0.2cm}

\ni are two of the eight Gell-Mann $3\times 3$ matrices (here, the approximation of $s_{\rm atm} = \pm 1$ {\it i.e.}, $c_{23} = 1/\sqrt2 = \pm s_{23}$ is experimentally satisfactory). Note that ${\footnotesize \frac{1}{2}}({\bf 1}^{(3)} + \varphi_3) = -{\footnotesize \frac{1}{2}}(\varphi_1 + \varphi_2)$. 

With the matrix $\varphi_3 $ as given in the third Eq. (11), it is not difficult to show from Eq. (18) that in the case of $s_{13} = 0$ the effective neutrino mass matrix $M$ is {\it invariant} under the third ($a = 3$) neutrino transformation (10),

\begin{equation}
\varphi_3 M \varphi_3 = M \,,
\end{equation}

\ni while for the first ($a = 1$) and second ($a = 2$) transformations (10) one gets

\begin{equation}
\varphi_{1,2} M \varphi_{1,2} = M + (m_1 - m_2) s_{\rm sol} (c_{23} \lambda_1 - s_{23} \lambda_4)
\stackrel{m_1-m_2 \rightarrow 0}{\longrightarrow} M 
\end{equation}

\ni {\it i.e.}, in the limit of $m_1-m_2 \rightarrow 0$ the effective mass matrix $M$ with $s_{13} =0$ is {\it invariant} also under the first and second transformations (10). So, in this limit, the matrix $M$ with $s_{13} =0$ is invariant under the whole four-group.

It is also not difficult to demonstrate that, inversely, the invariance (20), if imposed on $M$, implies for $s_{13}$ the restriction $s_{13} = 0$. In fact, Eqs. (6) for $M_{\alpha \beta}$ valid for generic $s_{13}$, when substituted into Eq. (20), lead {\it e.g.} to the equality

\begin{eqnarray}
M_{e \mu} & = & \left(\varphi_3 M \varphi_3 \right)_{e \mu} = c_{\rm atm}M_{e \mu} - s_{\rm atm}M_{e \tau} 
\nonumber \\ 
& = & M_{e \mu} +2 s_{23} c_{13}  s_{13} \left(c^2_{12}m_1 +s^2_{12} m_2 - m_3 \right) \,.
\end{eqnarray}

\ni This implies that $s_{13} = 0$ since $c_{13} \neq 0$. Then, with $c^2_{13} = 1$ the matrix $M$ must have the form (7) or (18), while with $c_{13} = 1$ the matrix $U$ has to be reduced to the form (2).

The proof that the restriction $s_{13} = 0$ follows from the invariance of $M$ described essentially by Eq. (20) (even if $c_{\rm atm} = \cos 2\theta_{23} \neq 0$) was presented previously in Ref. [12]. Such an invariance (with $c_{\rm atm} = \cos 2\theta_{23} = 0$ and $s_{13} = 0$) was considered also in Refs. [13] and (14) as well as in Ref. [10].

\vspace{0.2cm}

\ni {\bf 3. Duality of atmospheric and solar mixing angle in the case of $s_{13} = 0$}

\vspace{0.2cm}

Four $3\times 3$ matrices ${\bf 1}^{(3)},\,\mu_1\,,\,\mu_2\,,\, \mu_3$, where $\mu_a$ are defined by the unitary transformations 

\begin{equation}
\mu_a \equiv U^\dagger \varphi_a U \;\;(a = 1,2,3) \,, 
\end{equation}

\ni constitute in the case of $s_{13} = 0$ another reducible representation $ \underline{3} = \underline{2} + \underline{1}$ of the four-group, that is unitarily isomorphic to its previous representation  $ \underline{3} = \underline{1} + \underline{2} $, consisting of $3\times 3$ matrices ${\bf 1}^{(3)},\,\varphi_1\,,\,\varphi_2\,,\, \varphi_3$ introduced in Eqs. (12) and (11). In fact, with the use of Eqs. (11) for $\varphi_a$ (with any $c_{\rm atm}$ and $s_{\rm atm}$) and the form (2) of $U$ valid in the case of $s_{13} = 0$, we obtain

\begin{eqnarray}
\mu_1 & = &  \left(\begin{array}{ccc} -c_{\rm sol} & -s_{\rm sol} & 0 \\ -s_{\rm sol} & c_{\rm sol} & 0 \\ 0 & 0 & -1 \end{array}\right) = {\rm diag}(-c_{\rm sol}\sigma_3 - s_{\rm sol}\sigma_1,-1)  \,, \nonumber \\ 
\mu_2 & = & \left(\begin{array}{ccc} c_{\rm sol} & s_{\rm sol} & 0 \\ s_{\rm sol} & -c_{\rm sol} & 0 \\ 0 & 0 & -1 \end{array}\right) = {\rm diag}(c_{\rm sol}\sigma_3 + s_{\rm sol}\sigma_1,-1)  \,, \nonumber \\
\mu_3 & = & \left(\begin{array}{ccc} -1 & 0 & 0 \\ 0 & -1 & 0 \\ 0 & 0 & 1 \end{array}\right) = {\rm diag}(-{\bf 1}^{(2)},1)  \,. 
\end{eqnarray}

\vspace{0.2cm}

\ni Recall that $c_{\rm sol} =  c^2_{12} - s^2_{12} = \cos 2\theta_{12}$ and $s_{\rm sol} =  2c_{12} s_{12} = \sin 2\theta_{12}$. Here, it is convenient to write ${\bf 1}^{(3)} = {\rm diag} ({\bf 1}^{(2)}\,,\,1)$. Evidently, the four matrices ${\bf 1}^{(3)},\,\mu_1\,,\,\mu_2\,,\, \mu_3$ satisfy for any $c_{\rm sol}$ and $s_{\rm sol}$ the algebraic relations identical in form with Eqs. (13),

\begin{equation}
\mu_1 \mu_2 = \mu_3\;\,{\rm (cyclic)} \;\,\,,\;\,\, \mu^2_a= {\bf 1}^{(3)} \;\,,\;\, \mu_a \mu_b = \mu_b \mu_a
\,, 
\end{equation}

\ni and also the constraint identical in form with Eq. (14)

\begin{equation}
{\bf 1}^{(3)} + \mu_1+ \mu_2 + \mu_3 = 0 \,.
\end{equation}

\ni The matrices $\mu_a$ were already considered in Ref. [10], but in the formal limit of $s_{\rm sol} \rightarrow 1$ (in contrast to $s_{\rm atm} = \pm 1$ {\it i.e.}, $c_{23} = 1/\sqrt2 = \pm s_{23}$, the approximation $s_{\rm sol} = \pm 1$ {\it i.e.}, $c_{12} = 1/\sqrt2 = \pm s_{12}$ is experimentally not satisfactory).

>From Eqs. (23), (1) and (10) we can infer that 

\begin{equation}
\left(\begin{array}{l} \nu'_1 \\ \nu'_2 \\ \nu'_3 \end{array}\right)_{\!\!a} = \mu_a \left(\begin{array}{l} \nu_1 \\ \nu_2 \\ \nu_3 \end{array}\right) = U^\dagger \varphi_a \left(\begin{array}{l} \nu_e \\ \nu_\mu \\ \nu_\tau \end{array}\right) = U^\dagger \left(\begin{array}{l} \nu'_e \\ \nu'_\mu \\ \nu'_\tau \end{array}\right)_{\!\!a}\;\,(a = 1,2,3) \,.
\end{equation}

\vspace{0.2cm}

\ni Thus, the four-group transformations (27) of mass neutrinos $\nu_i $ (produced by three matrices $\mu_a$) are {\it covariant} under the neutrino mixing (1): they transit into the four-group transformations (10) of flavor neutrinos $\nu_\alpha$ (generated by three matrices $\varphi_a$).

In addition, Eqs (11) and (24) involving $\theta_{\rm atm} = 2\theta_{23}$ and $\theta_{\rm sol} = 2\theta_{12}$, respectively, being related through the unitary transformations (23), tell us that the atmospheric and solar mixing angles, $\theta_{\rm atm} = 2\theta_{23}$ and $\theta_{\rm sol} = 2\theta_{12}$ are in a way mutually {\it dual} in the process of neutrino mixing described in Eq. (1),

\begin{equation}
\left(\begin{array}{l} \nu_1 \\ \nu_2 \\ \nu_3 \end{array}\right) = U^\dagger \left(\begin{array}{l} \nu_e \\ \nu_\mu \\ \nu_\tau \end{array}\right) \,. 
\end{equation}

\vspace{0.2cm}

\ni Beside the duality relations $U^\dagger \varphi_a \left(c_{\rm atm}, s_{\rm atm} \right) U = \mu_a \left(c_{\rm sol}, s_{\rm sol} \right)$ with $\varphi_a$ and $\mu_a$  given in Eqs. (11) and (24), respectively, we can show that

\begin{equation}
U^\dagger\left(c_{23} \lambda_1- s_{23} \lambda_4 \right) U = c_{12} \lambda_1 - s_{12} \lambda_3 \,, 
\end{equation}

\ni where

\begin{equation}
\lambda_3 = \left( \begin{array}{rrr} 1 & 0 & 0 \\ 0 & -1 & 0  \\ 0 & 0 & 0 \end{array} \right) 
\end{equation}

\vspace{0.2cm}

\ni is the third Gell-Mann $3\times 3$ matrix. The duality relation (29) follows from a direct calculation using the form (2) of $U$ valid in the case of $s_{13} = 0$.

The formula (27) compared with Eq. (28) shows that the effective neutrino mixing matrix $U$ transforming $\nu_i$ into $\nu_\alpha$ is {\it invariant} under the four-group. The same conclusion follows as a tautology from Eqs. (23) rewritten in the equivalent form

\begin{equation}
\varphi_a U \mu_a = U \;\; (a = 1,2,3) \,,
\end{equation}

\ni where $\varphi_a$ and $\mu_a$ belong to two unitarily isomorphic representations $ \underline{3} $ of the four-group.

\vspace{0.3cm}

{\bf 4. Conclusion}

\vspace{0.2cm}

Thus, if and only if $s_{13} = 0$, the generic form of the effective neutrino mass matrix $M$ becomes invariant under the subgroup $Z_2$ of the four-group $Z_2\times Z_2$ represented by ${\bf 1^{(3)}}$ and $\varphi_3$. In the approximation of $m_1 = m_2$, the matrix $M$ becomes invariant under the whole four-group represented by the matrices ${\bf 1}^{(3)},\,\varphi_1\,,\,\varphi_2\,,\, \varphi_3$.

In the case of $s_{13} = 0$, the atmospheric and solar mixing angles, $\theta_{\rm atm} = 2\theta_{23}$ and $\theta_{\rm sol} = 2\theta_{12}$, turn out to be mutually dual in the process of neutrino mixing, what means that $U^\dagger \varphi_a(c_{\rm atm}, s_{\rm atm}) U = \mu_a(c_{\rm sol}, s_{\rm sol})\; (a = 1,2,3)$, where $c_{\rm atm} = \cos \theta_{\rm atm}$, $s_{\rm atm} = \sin \theta_{\rm atm}$ and $c_{\rm sol} = \cos \theta_{\rm sol}$, $s_{\rm sol} = \sin \theta_{\rm sol}$. Here, ${\bf 1}^{(3)},\,\varphi_1\,,\,\varphi_2\,,\, \varphi_3$ and ${\bf 1}^{(3)},\,\mu_1\,,\,\mu_2\,,\, \mu_3$ constitute two unitarily isomorphic reducible representations $\underline{3}$ of the four-group (producing four-group transformations of three flavor and three mass neutrinos, respectively).

\vfill\eject

~~~~

\vspace{0.4cm}

{\centerline{\bf References}}

\vspace{0.4cm}

{\everypar={\hangindent=0.7truecm}
\parindent=0pt\frenchspacing

{\everypar={\hangindent=0.7truecm}
\parindent=0pt\frenchspacing

~[1]~Q.R. Ahmad {\it et al.} (SNO Collaboration), {\it Phys. Rev. Lett.} {\bf 87}, 071301 (2001); {\tt nucl--ex/0309004}.

\vspace{2.0mm}

~[2]~Y. Fukuda {\it et al.} (SuperKamiokande Collaboration), {\it Phys. Rev. Lett.} {\bf 81}, 1562 (1998); {\it Phys. Lett.} {\bf B 467}, 185 (1999).

\vspace{2.0mm}

~[3]~M.H. Ahn {\it et al.} (K2K Collaboration), {\it Phys. Rev. Lett.} {\bf 90}, 041801 (2003).

\vspace{2.0mm}

~[4]~K. Eguchi {\it et al.} (KamLAND Collaboration), {\it Phys. Rev. Lett.} {\bf 90}, 021802 (2003).

\vspace{2.0mm}

~[5]~For a recent review {\it cf.} V. Barger, D. Marfatia, K. Whisnant, {\it Int. J. Mod. Phys.} {\bf E 12}, 569 (2003); M.~Maltoni {\it et al.}, {\tt hep--ph/0405172}; M.C.~Gonzalez-Garcia, {\tt hep--ph/0410030}; G.~Altarelli {\tt hep--ph/0410101}.

\vspace{2.0mm}

~[6]~M. Apollonio {\it et al.} (Chooz Collaboration), {\it Eur. Phys. J.} {\bf C 27}, 331 (2003).

\vspace{2.0mm}

~[7]~C. Athanassopoulos {\it et al.} (LSND Collaboration), {\it Phys. Rev. Lett.} {\bf 77}, 3082 (1996); {\it Phys. Rev. } {\bf C 58}, 2489 (1998); A. Aguilar {\it et al.}, {\it Phys. Rev.} {\bf D 64}, 112007 (2001).

\vspace{2.0mm}

~[8]~H. Murayama, T. Yanagida, {\it Phys. Rev.} {\bf B 52}, 263 (2001); G. Borenboim, L.~Borissov, J. Lykken, A.Y. Smirnov, {\it J.High Energy Phys.} {\bf 0210}, 001 (2002); G. Borenboim, L. Borissov,  J. Lykken, {\tt hep--ph/0212116}. 

\vspace{2.0mm}

~[9]~A.O. Bazarko {\it et al.}, {\tt hep--ex/9906003}.

\vspace{2.0mm}

[10]~W. Kr\'{o}likowski, {\it Acta Phys. Pol.} {\bf B 34}, 4125 (2003); {\it Acta Phys. Pol.} {\bf B 34}, 4157 (2003). 

\vspace{2.0mm}

[11]~E.P. Wigner, {\it Group theory and its application to the quantum mechanics of atomic spectra}, Academic Press, New York and London, 1959.

\vspace{2.0mm}

[12]~W. Grimus, A.S. Yoshipura,S. Kaneko, L. Lavoura, H. Savanaka, M. Tanimoto, {\tt hep--ph/0408123}; and references therein.

\vspace{2.0mm}

[13]~W. Grimus, L. Lavoura, {\it Acta Phys. Pol.}, {\bf B 34}, 5393 (2003); and references therein.

\vspace{2.0mm}

[14]~E. Ma, {\tt hep--ph/0307016}; {\tt hep--ph/0409075}; and references therein.

\vfill\eject

\end{document}